**The Measurement of a Temperature Dependence of the Count Rate of Single Electrons Emitted From Copper by Means of a Multicathode Counter.**

Anatoly Kopylov, Igor Orekhov, Valery Petukhov

Institute for Nuclear Research of Russian Academy of Sciences,

117312 Moscow, Prospect of 60th Anniversary of October Revolution 7A, Russia

beril@inr.ru

ABSTRACT.

It was found that at cryogenic temperatures the spontaneous emission of single electrons from cathode of PMT is increased by lowering the temperature of PMT. This effect hasn't found yet a satisfactory explanation. It was suggested by us that the source of "cryogenic" dark current is a conversion of hidden photons on a metallic surface. To check this hypothesis a series of measurements have been performed using a multicathode counter developed by us as a detector of single electrons. Upon the results of measurements during 78 days in configurations $R_1$ иад $R_2$ of a multicathode counter with a copper cathode the dependence of the counting rate of single electrons on temperature has been obtained. The results obtained are in agreement with the hypotheses that hidden photons are the source of a "cryogenic" dark current. These very preliminary results have to be checked by further measurements.

In the work [1, 2] the effect has been found of the increase of dark rate of single electrons emitted from a surface of PMT at lowering the temperature of PMT from -20°C till a cryogenic one. This effect has been observed both with bi-alkali (K-Cs-Sb) and with tri-alkali (Sb-Na-K-Cs) cathodes, with a platinum backing and without it. By observing this effect it was found also that electrons are emitted in bursts and the emission is somewhat non statistical in character. The bursts itself had a statistical distribution in time but the events inside bursts were correlated. This character of electron emission was suggesting that the traps on a surface of a cathode are responsible for the effect. The energy from external source gets accumulated in traps and upon reaching of a certain threshold value it gets released by a sequence of single electrons of an avalanche type. The time correlation has been observed between pulses in an avalanche and the total energy accumulated by the trap. In a work [1] this effect has been observed by lowering the temperature till -70°C, in a work [2] – till 4°K. No satisfactory explanation has been suggested till nowadays. In a paper [3] we have considered the possibility to explain this effect by a conversion of hidden photons on a surface of a cathode of PMT. Hidden photons were first considered by L.B.Okun' [4] as a modification of electrodynamics where the field of photons is of a spatially constant mode oscillating with frequency $\omega = m_{\gamma'}$, where $m_{\gamma'}$ – mass of a hidden photon. In this hypothesis it is supposed that the energy released during the conversion of hidden photon into a photon at the surface of a cathode gets absorbed by traps and by reaching a certain critical level it gets released in a burst of single electrons. A competing process is the emission of energy by thermal radiation. By lowering the temperature the role of a competing thermal radiation gets decreased and the rate of emission of single electrons gets increased. If this model is true at a temperature 4°K when the effect of thermal radiation can be neglected the rate of emission of single electrons should be inversely proportional to a work function of a material of cathode. If to extrapolate the spectrum of single electrons from a cathode of PMT, see Fig 16 in [5], a work function $\varphi_w$ for tri-alkali cathode is of about 0.4 eV. If to use copper as a material of a cathode with a work function of about 4 eV the dark count rate of single electrons at 4°K should be an order of magnitude lower. In a work [2] it was found that the rate of emission of single electrons from a cathode of PMT depends upon temperature at low temperatures as

$$I = G \cdot A \cdot \exp(-T/T_c) \qquad (1)$$

where I – a dark count rate, $G \approx 5$ cm$^{-2}$s$^{-1}$, $T_c \approx 100$ °K

Figure 1 shows a dark rate observed in experiment [2] as a function of temperature and also what should be observed for a opper cathode. The exponent in the expression (1) can be interpreted as a part of energy absorbed by a trap which gets released by the emission of single electrons. At absolute zero T = 0 °K we have $\exp(-T/T_c) = 1$ and the energy absorbed by traps gets completely released by the emission of single electrons. By the conversion of hidden photon the energy gets absorbed equal to a mass of a hidden photon: $\omega = m_{\gamma'}$. From here one can get at absolute zero that quantum yield for emission of single electron

$$\eta = m_\gamma / \varphi_w \qquad (2)$$

.

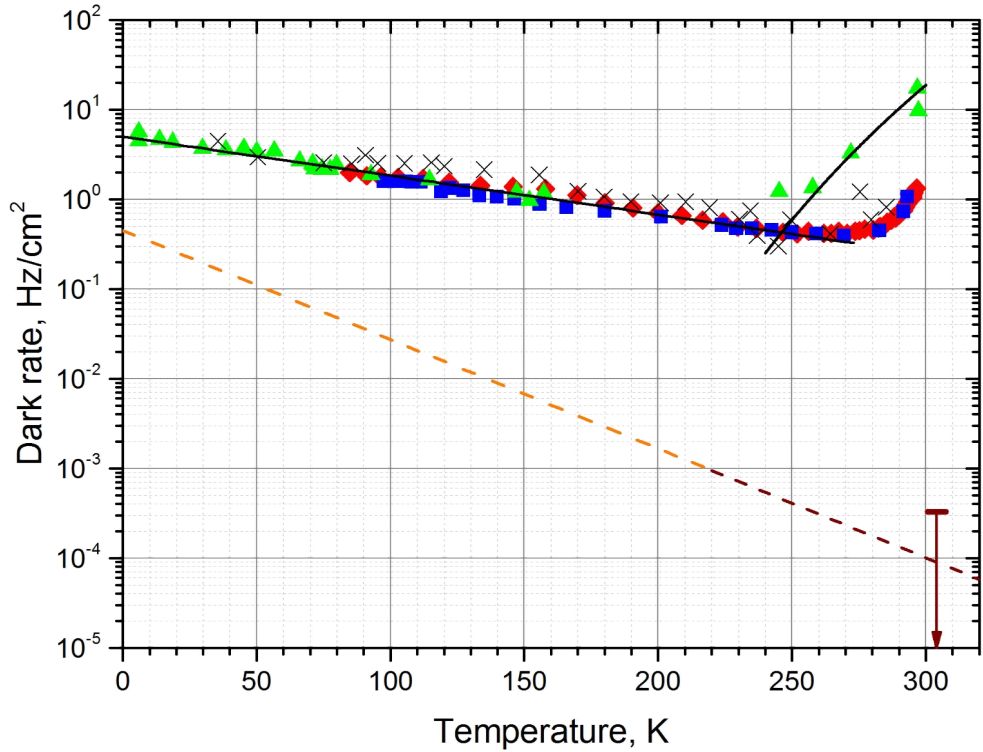

Figure.1 A tempearature dependence of dark rate of single electrons. Upper line – the data from [2], lower line – expected effect for a copper cathode. The upper limit at room temperature is taken from [7].

To register the effect from hidden photons the method of a multicathode counter has been developed by us [6, 7]. In this method it is found the difference of the count rates ($R_1 - R_2$) of single electrons in configurations 1 and 2 of a multicathode counter. In a first configuration cathode C3 is kept at a potential which enables a drift of electrons towards a central counter with a cathode C1where due to high coefficient of gas amplification ( >10⁵ ) single electrons are efficiently registered. In a second configuration a cathode C3 is kept at a potential which rejects electrons emitted from a copper cathode thus preventing them to drift towards a central counter.

A calibration of the counter has been performed by a source of X-rays $^{55}$Fe, and in a single electron mode by UV emission of a mercury lamp through e special quarts window. For anode of the counter a gold plated tungsten-rhenium wire of 20 μm in diameter. A signal from a wire is coming on the input of charge sensitive preamplifier Zaryad fabricated in Russia with a sensitivity 0.5 V/pC and from the output of it to a digitizer NI-5152 8 bit.

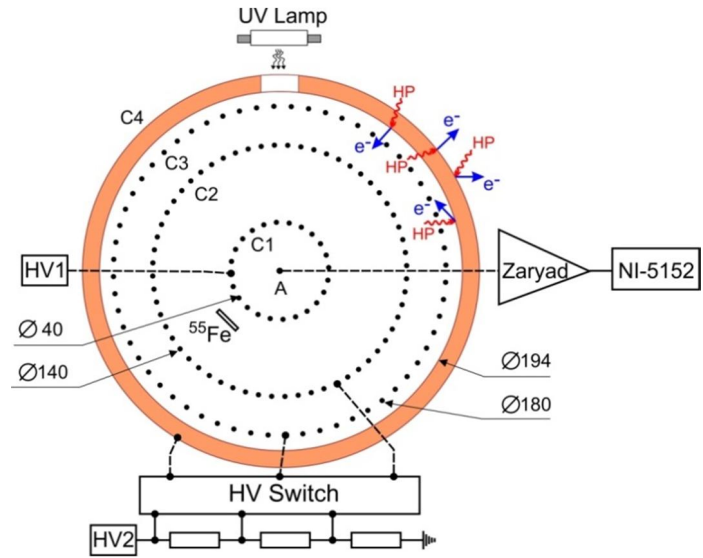

Figure 2. A simplified scheme of a multicathode counter.

The effect from hidden photons is found as

$$R_{MCC} = (R_1 - R_2)/\varepsilon \qquad (3)$$

here: ε – is the efficiency of the counting of single electrons determined from the results of calibration by a mercury lamp. The design of the counter and the technique developed are described in more details in [6, 7]. Sensitivity of a multicathode counter is described by the expression [7]

$$\chi_{sens} = 2.9 \times 10^{-12} \left( \frac{R_{MCC}}{\eta\ 1\ \text{Hz}} \right)^{\frac{1}{2}} \left( \frac{m_{\gamma'}}{1\ \text{eV}} \right)^{\frac{1}{2}} \left( \frac{0.3\ GeV/cm^3}{\rho_{CDM}} \right)^{\frac{1}{2}} \left( \frac{1\ \text{m}^2}{A_{MCC}} \right)^{\frac{1}{2}} \left( \frac{\sqrt{2/3}}{\alpha} \right) \qquad (4)$$

Here: $A_{MCC}$ – a surface of a cathode, α – a parameter of anisotropy which is equal to $\sqrt{2/3}$ for isotropic flux, η – quantum efficiency for a given energy of a photon. If to put in this expression a quantum efficiency from (2) and instead of $R_{MCC}$ – the rate of single electrons I = G·A at 4°K then we will get an expression independent on a mass of a hidden photon

$$\chi = 2.9 \times 10^{-10} \left( G \cdot \varphi_W \right)^{1/2} \left( \frac{0.3 \ GeV/cm^3}{\rho_{CDM}} \right)^{1/2} \left( \frac{\sqrt{2/3}}{\alpha} \right) \tag{5}$$

If to put in this expression a measured [2] at 4 °K dark rate $G \approx 5$ cm$^{-2}$s$^{-1}$ and to take a work function $\varphi_w = 0.4$ eV, then one can find a mixing angle at $\rho_{CDM} = 0.3$ GeV/cm$^3$ and $\alpha = \sqrt{2/3}$

$$\chi = 4.1 \cdot 10^{-10} \tag{6}$$

which, as one can see, is of a rather large value. Which would be a mass of a hidden photon, this is still an open question. If to take limitations from [8, 9] it is less than 2 eV. To check a hypothesis that hidden photons are really the source of a "cryogenic" dark rate the measurement of the rates of single electrons emitted from a copper cathode have been performed at temperatures 26, 31 and 36°C. The measurements have been conducted during 78 days and more than 15 TB of information have been collected as a result of digitizing the shapes of the pulses from charge sensitive preamplifier. The analysis of the data has been conducted in off line, the useful signal considered to be within interval from 3 to 30 mV with a short leading edge corresponding to a time of a drift of positive ions to cathode and a slow tail, corresponding to a baseline restoration of charge sensitive preamplifiers and with a 'good' prehistory of the events corresponding to small deviation of a baseline from zero potential. Upon the results of data treatment it was found that $r_{MCC}(26°C) = (0.98 \pm 0.22) \cdot 10^{-4}$ Hz·cm$^{-2}$, $r_{MCC}(31°C) = (0.75 \pm 0.15)$ $\cdot 10^{-4}$ Hz·cm$^{-2}$, $r_{MCC}(36°C) = (0.69 \pm 0.23) \cdot 10^{-4}$ Hz·cm$^{-2}$. Here $r_{MCC}$ means count rate in Hz per cm$^2$. The averaged by all three temperatures value was found to be $r_{MCC} = (0.79 \pm 0.11) \cdot 10^{-4}$ Hz·cm$^{-2}$. As one can see it is above zero at more than 7 standard deviations. The falling with temperature dependence proves that the effect of thermionic emission of single electrons can be neglected. The measured time dependence is presented at Fig. 3. One can see that these points are in a reasonable agreement with the expected effect if this hypothesis is true. One can see also that our method has high sensitivity to measure the effect 3 ÷ 4 orders of magnitude lower than demonstrated in [2].

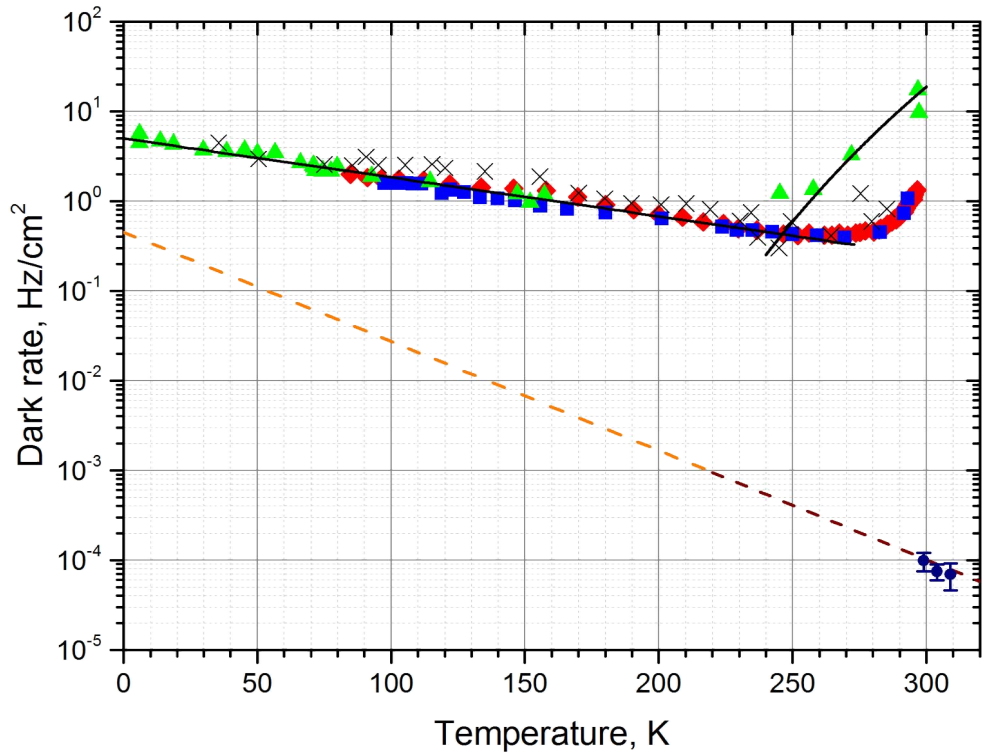

Figure 3. A temperature dependence of a dark rate of counting of single electrons. . Upper line – the data from [2], lower line – expected effect for a copper cathode. The points are the results of measurements at three temperatures: 26, 31 and 36°C.

Certainly, these results are very preliminary and should be checked in further measurements at lower temperatures. Because we use in the counter P10 gas, i.e. argon plus 10% methane mixture, the minimal temperature which can be used for this mixture is about 200-220 °K. This range is depicted at Fig 3 by more dark dashed line. We are planning to perform the measurements in this range. The results obtained in this work can be considered only as a hint of possible physical process. They need a further clarification.

In a conclusion the authors express deep gratitude to E.P.Petrov for active participation and good help in fabrication of the counter and also to all members of a machine shop facility of INR RAS where all mechanical and welding works have been conducted.